\documentclass[aps,prl,showpacs,twocolumn,superscriptaddress]{revtex4}
\usepackage{amssymb}
\usepackage{graphicx}% Include figure files
\usepackage{dcolumn}% Align table columns on decimal point
\usepackage{bm}% bold math
\usepackage{color}
\begin{document}

\title{Phonon induced Rabi frequency renormalization of optically driven single InGaAs/GaAs quantum dots}

\author{A.~J.~Ramsay}
\email{a.j.ramsay@shef.ac.uk}
\affiliation{Department of Physics
and Astronomy, University of Sheffield, Sheffield, S3 7RH, United
Kingdom}

\author{T.~M.~Godden}
\affiliation{Department of Physics
and Astronomy, University of Sheffield, Sheffield, S3 7RH, United
Kingdom}

\author{S.~J.~Boyle}
\affiliation{Department of Physics
and Astronomy, University of Sheffield, Sheffield, S3 7RH, United
Kingdom}

%\author{Achanta~Venu~Gopal}
%\affiliation{DCMP \& MS, Tata Institute of Fundamental Research, Mumbai 400 005, India}

\author{E.~M.~Gauger}
\affiliation{Department of Materials, University of Oxford, Oxford OX1 3PH, United Kingdom}

\author{A.~Nazir}
\affiliation{Department of Physics and Astronomy, University College London, London, WC1E 6BT, United Kingdom}

\author{B.~W.~Lovett}
\affiliation{Department of Materials, University of Oxford, Oxford OX1 3PH, United Kingdom}
\affiliation{School of Engineering and Physical Sciences, Heriot-Watt University, Edinburgh EH14 4AS, United Kingdom}

\author{A.~M.~Fox}
\affiliation{Department of Physics and Astronomy, University of
Sheffield, Sheffield, S3 7RH, United Kingdom}

\author{M.~S.~Skolnick}
\affiliation{Department of Physics and Astronomy, University of
Sheffield, Sheffield, S3 7RH, United Kingdom}

\date{\today}% It is always \today, today,
             %  but any date may be explicitly specified

\begin{abstract}
We study optically driven Rabi rotations of a quantum dot exciton transition between 5 and 50~K, and for pulse-areas of up to $14\pi$. In a high driving field regime, the decay of the Rabi rotations is nonmonotonic, and the period decreases with pulse-area and increases with temperature. By comparing the experiments to a weak-coupling model of the exciton-phonon interaction, we demonstrate that the observed renormalization of the Rabi frequency is induced by fluctuations in the bath of longitudinal acoustic phonons, an effect that is a phonon analogy of the Lamb-shift.
\end{abstract}
% PACS, the Physics and Astronomy
                             % Classification Scheme.
%\keywords{Suggested keywords}%Use showkeys class option if keyword
                              %display desired
\maketitle

The decoherence of a two-level system or qubit in the solid-state is often understood in terms of its interaction with a reservoir of bosons \cite{Leggett_rmp} or half-integer spins \cite{Hanson_sci}. An important issue with respect to both decoherence physics and the performance of quantum logic operations is how these interactions are modified by a driving field. In the case of bosons, this issue has received considerable theoretical attention \cite{Leggett_rmp}, but as far as we are aware, the only experimental studies to support this work are restricted to the field of superconducting qubits \cite{Wilson_prl,Wilson_prb}. Here we use Rabi rotation measurements of quantum dot excitons \cite{Stievater_prl,Zrenner_nat,Wang_prb} to study the dephasing of a solid-state driven two-level system coupled to a boson bath, where recently we identified longitudinal acoustic phonons as the principal cause of the intensity damping of optically driven Rabi rotations in InAs/GaAs quantum dots \cite{Ramsay_prl2010}.

In this letter, we report experimental evidence for a phonon-induced shift in the Rabi frequency of an optically-driven excitonic transition in a semiconductor quantum dot. A single laser pulse drives a Rabi rotation between the ground and neutral exciton states of a single InGaAs/GaAs quantum dot, where the pulse-area, the time-integral of the Rabi frequency, is controlled via the incident power. Compared to a two-level atom picture, the resulting Rabi rotations are modified by the exciton-phonon interaction. We observe key signatures of the dynamics of a driven two-level system coupled to a boson bath.  Firstly, the period of the Rabi rotations decreases with the pulse-area, and increases with temperature implying a renormalization of the Rabi frequency as theoretically anticipated in refs. \cite{Forstner_prl,Machnikowski_prb,Krugel_apb,Vagov_prl,Nazir_prb}. %For temperatures of up to 35~K, the data is well described by a weak-coupling model of the exciton-phonon interaction \cite{Nazir_prb}.
This shift in the Rabi energy is a result of fluctuations in the phonon bath, and is an effect analogous to the Lamb-shift observed in the spectra of atomic hydrogen. Secondly, the damping of the Rabi rotations is nonmonotonic and exhibits a roll-off behavior as the Rabi energy starts to exceed the bandwidth $\hbar\omega_c$ of the exciton-phonon coupling.

Since our previous work \cite{Ramsay_prl2010}, we have greatly improved our experimental setup by swapping a cold-finger cryostat for a helium bath cryostat with the sample mounted on low temperature piezoelectric stages. This provides access to lower temperatures (4.2~K vs 15~K), higher mechanical stability, and enables the use of a shorter focal length objective. Overall this increases the range of the pulse-area from $6\pi$ to $14\pi$, providing the bandwidth in the Rabi energy needed to resolve the new physics reported here. The experiments are performed on a single quantum dot embedded in
an n-i-Schottky diode structure. Uncapped test specimens exhibit an ensemble of InGaAs/GaAs dots with heights of 3-4 nm, and base diameters of 25-30 nm. %Based on cross-sectional STM measurements on similar samples, \cite{Bruls_apl}
%we expect the dots to have a truncated pyramid shape, with an Indium composition gradient varying linearly from $\sim\mathrm{InAs}$ at the top, to $\sim\mathrm{In_{0.5}Ga_{0.5}As}$ at the base.
%At $T\sim 15~K$, the emission energies of the ensemble peak at about 975-nm, with a full-width half-maximum (FWHM) of $\sim 30~\mathrm{nm}$.
The measurements presented here are from a single dot emitting at 951~nm  in the high energy tail of the dot distribution, and energetically distant from the wetting layer that emits at 861 nm. The biexciton binding energy is $1.9~\mathrm{meV}$. The dot is excited with a single laser pulse, and the final occupation of the exciton state is measured using photocurrent detection \cite{Zrenner_nat}. A background photocurrent proportional to the incident power is subtracted from all data, attributed to absorption of scattered light by other dots in the sample \cite{Stufler_prb2}. Further details of the sample, and the optical system can be found in ref. \cite{Boyle_prb}.

\begin{figure}
\begin{center}
\includegraphics[scale=2.4]{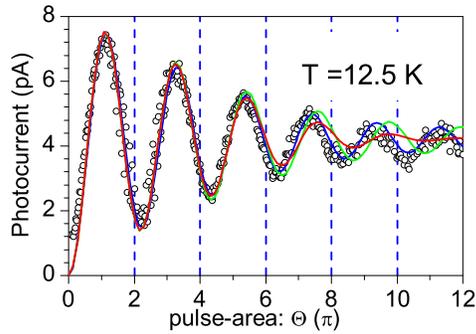}
\end{center}
\caption{($\circ$)A Rabi rotation measurement at $12.5~\mathrm{K}$. The full lines are calculated using Eqs. (\ref{eqn:Bloch2},\ref{eqn:Bloch3}).
(red) ($K(\Omega)$ is real, and $\mathcal{P}=1$)
In a low driving field regime ($\Theta<6\pi$), the data are described by an $\Omega^2$ damping, but at higher fields this model overdamps the oscillation. (green) ($K(\Omega)$ is real, $\mathcal{P}(\Omega)= e^{-\Omega^2/\omega_c^2}$) The roll-off in the damping is reproduced by including the frequency dependence of the exciton-phonon form-factor $\mathcal{P}(\Omega)$. With a constant rotation frequency, this model provides a `clock-signal' against which the nonlinearity of the rotation angle with $\Theta$ is clearly observed. (blue)($K(\Omega)$ is complex) By including the imaginary part of $K(\Omega)$ the  variation of rotation angle with $\Theta$ is reproduced.
}\label{fig:fig1}
\end{figure}

Figure \ref{fig:fig1} presents a  Rabi rotation measurement at a temperature of $12.5~\mathrm{K}$, where the qualitative features of the data are clearly expressed. A single laser pulse with a Gaussian envelope resonantly excites the neutral exciton transition. Circular polarization and a narrow 0.2-meV FWHM spectral width are used to suppress excitation of the two-photon biexciton transition \cite{Stufler_prb}, and of multi-excitons \cite{Patton_prl}. The bare Rabi frequency $\Omega(t)$ of the laser pulse is described by  $\Omega(t)=\frac{\Theta}{2\tau\sqrt{\pi}}\exp{(-(\frac{t}{2\tau})^2)}$, where the pulse-duration $\tau = 4~\mathrm{ps}$, and the pulse-area $\Theta=\int_{-\infty}^{\infty}\Omega(t)dt$. The photocurrent $PC$, which is proportional to the final occupation of the exciton state following resonant excitation, is measured as a function of the square-root of the incident power, which is proportional to the pulse-area $\Theta$. The photocurrent exhibits a damped oscillation with pulse-area. To help illustrate the main qualitative features of the data, calculations of various models of the driven dephasing are presented in fig. \ref{fig:fig1} as color-coded lines. The traces are scaled to fit the data at low pulse-area and will be discussed shortly.

To analyze the data, we use a model described by a pair of Bloch-equations:
\begin{eqnarray}
\dot{s}_y = \{\Omega+\Im[K(\Omega,T)]\}s_z-\{\Gamma_2^*+\Re[K(\Omega,T)]\}s_y \label{eqn:Bloch2}\\
\dot{s}_z = -\Omega s_y \label{eqn:Bloch3}
\end{eqnarray}
\noindent where $\mathbf{s}=(s_x,s_y,s_z)$ is the Bloch-vector of the two-level system in the rotating-frame of the laser, composed of the crystal ground, and the exciton states. $s_z$ represents the population inversion, and $s_y$ the imaginary component of the excitonic dipole. The photocurrent is a measure of the final occupation of the exciton state: $PC\propto (1+s_z)/2$. In the case of on-resonance excitation, where the initial state is $\mathbf{s}=(0,0,-1)$, as discussed here, the $s_x$ equation is decoupled from Eqs. (\ref{eqn:Bloch2},\ref{eqn:Bloch3}) and does not influence the dynamics of the measured $s_z$-basis. %$\Omega$ is the Rabi frequency, and $T$ the temperature.
 $\Gamma_2^*$ is a phenomenological dephasing rate used to account for electron-tunneling, and other possible sources of driving-field independent dephasing. The complex response function $K(\Omega,T)$ describes the Rabi-frequency dependence of the exciton-phonon interaction. The real part $\Re$ gives rise to a rate of dephasing, and the imaginary part $\Im$ shifts the rotation frequency of the Bloch-vector; $\Re$ and $\Im$ satisfy a Kramers-Kronig relationship.

Eqs. (\ref{eqn:Bloch2},\ref{eqn:Bloch3}) are derived using a weak-coupling Born-Markov approximation \cite{Nazir_prb} that treats the exciton-phonon interaction Hamiltonian
$H_{I}= \vert X\rangle\langle X\vert \sum_\mathbf{q}\hbar(g_ \mathbf{q}b^{\dagger}_ \mathbf{q}+g_ \mathbf{q}^*b_ \mathbf{q})$
as a perturbation to second order, where $b_{\mathbf{q}}^{\dagger}$, $b_{\mathbf{q}}$ are the creation and annihilation operators for a longitudinal acoustic phonon of wave-vector $\mathbf{q}$. The material dependent coupling strengths $g_{\mathbf{q}}$ are given in ref. \cite{Krummheuer_prb}. %by
%$g_{\mathbf{q}}=\frac{q(D_e  \mathcal{P}[\psi^{e}(\mathbf{r})] -D_h  \mathcal{P}[\psi^{h}(\mathbf{r})])}{\sqrt{2\mu\hbar\omega_{\mathbf{q}}V}}$
 %\cite{Krummheuer_prb} where $\mu$ is the mass density of the host material, $V$ the lattice volume, $D_{e(h)}$ the respective bulk electron (hole) deformation potential coupling constant, and $\mathcal{P}[\psi^{e(h)}]$ denotes the form factor of the electron (hole) wavefunction.
 $K(\Omega,T)=\int_0^{\infty}dt~e^{i\Omega t}\tilde{K}(t,T)$, where $\tilde{K}(t,T)$ is the time-domain response of the exciton-phonon system:
\begin{equation}
\tilde{K}(t,T) = \int_0^{\infty} d\omega J(\omega) \mathrm{coth}(\frac{\hbar\omega}{2k_BT})\cos{\omega t},
\end{equation}
\noindent where $J(\omega)=\sum_{\mathbf{q}}\vert g_{\mathbf{q}}\vert^2\delta(\omega-\omega_{\mathbf{q}}) $ is the spectral density of the exciton-phonon interaction, and can be approximated by
$J(\omega)\approx  \frac{\hbar A}{\pi k_B}\omega^3 \mathcal{P}^2(\omega)$. $A$ is defined such that $\lim_{\Omega\rightarrow 0}\Re{[K]}=AT\Omega^2$ \cite{Ramsay_prl2010}. The form-factor, which is the Fourier transform of the probability distribution of the carriers, is approximated as $\mathcal{P}(\omega)\approx e^{-\omega^2/2\omega_c^2}$ \cite{Nazir_prb}.

A physical interpretation of the model is as follows. The laser couples the crystal-ground and exciton states to form optically dressed states separated by a Rabi splitting. Emission and absorption of LA-phonons with an energy equal to the Rabi-splitting results in a relaxation between the dressed states that leads to a rate of dephasing that is a function of the instantaneous Rabi frequency. To satisfy a Kramers-Kronig relationship, the dephasing is accompanied by a shift to the Rabi splitting induced by  fluctuations of the LA-phonon bath.

\begin{figure}
\begin{center}
\includegraphics[scale=2]{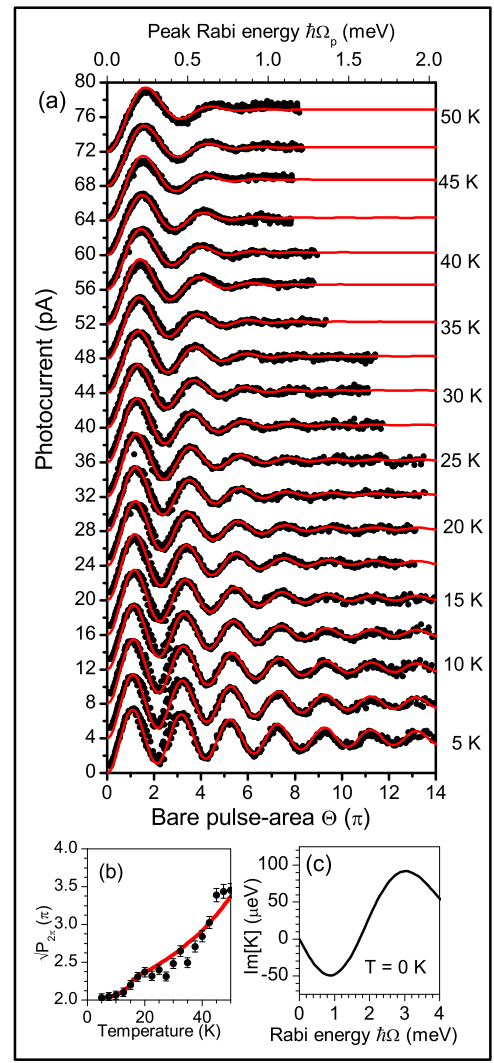}
\end{center}
\caption{(a) Temperature dependence of the Rabi rotations. Red-lines are calculated using $A=11.2~\mathrm{fs.K^{-1}}$,  $\hbar\omega_c=1.44~\mathrm{meV}$, and $\Gamma_2^*=0.025~\mathrm{ps^{-1}}$, where $A$ and $\hbar\omega_c$ are extracted from a fit to all data. The bottom axis is scaled to the bare pulse-area $\Theta$, while the top axis indicates the peak Rabi energy of the laser pulse.  (b) Square-root of the incident power $\sqrt{P}_{2\pi}$ corresponding to the first minimum plotted vs temperature. $\sqrt{P}_{2\pi}$ is scaled to the pulse-area $\Theta$ for the case of $T=5~\mathrm{K}$. The trace is calculated assuming the scaling parameter $a=1$. (c) Calculation of the Lamb-shift, $\Im{[K]}$ at $T=0~\mathrm{K}$, as a function of Rabi energy.
}\label{fig:fig2}
\end{figure}

Now we return to fig. \ref{fig:fig1}, to consider the qualitative features of the Rabi rotations using the results of Eqs. (\ref{eqn:Bloch2},\ref{eqn:Bloch3}) to help identify these features. Values of $A=11.2~\mathrm{fs~K^{-1}}$, and $\hbar\omega_c=1.44~\mathrm{meV}$ have been used, found from fits to fig. \ref{fig:fig2}(a), discussed below.  (i) The red-trace is calculated by neglecting the imaginary part of $K$, and approximating the form-factor as $\mathcal{P}\approx 1$. The pulse-area used in the calculations is scaled to provide best-fits to the data. For low pulse-areas of $\Theta <6\pi$, the data are well described by the red-trace, which is consistent with our previous work \cite{Ramsay_prl2010}. In the low-driving field, high temperature regime ($\Omega <\omega_c$, $2k_BT\gg \hbar\Omega$) the rate of dephasing is $\Re{[K]}\approx AT\Omega^2$, which is the expected behavior of a two-level system coupled to a three-dimensional boson bath \cite{Leggett_rmp}. However at higher pulse-areas, the calculation is over damped, high-lighting the nonmonotonic character of the decay. (ii) The green-trace is calculated  by again neglecting the imaginary part of $K$, but now including the frequency dependence of the form-factor $\mathcal{P}(\Omega)$. In this way the envelope of the Rabi rotation is reproduced. Hence, the decay is explained by a rate of dephasing that exhibits a cut-off behavior, where for driving fields greater than the cut-off energy ($\hbar\Omega>\hbar\omega_c$), the phonon-bath can no longer follow the driving field, leading to a reduction in the rate of dephasing. This calculation assumes that the rotation frequency of the Bloch-vector is equal to the Rabi frequency, and thus provides a periodic-signal against which the nonlinear dependence of the Rabi rotation angle with pulse-area is clearly observed. (iii) The blue-trace is calculated using the full model. Now both the envelope and the variation of Rabi rotation angle with pulse-area $\Theta$  are reproduced, demonstrating that the nonlinear rotation angle results from the imaginary part of the {\it same exciton-phonon response} that describes the damping.

To further confirm that the nonmonotonic decay and the nonlinear rotation angle are a result of the  exciton-phonon interaction, a set of Rabi rotation measurements as a function of temperature \cite{footnote3} was performed.  The data are presented in
 fig. \ref{fig:fig2}(a). The Rabi rotations are offset for clarity, and plotted as a function of the pulse-area deduced from the fits. For each temperature, the laser is tuned to resonance with the exciton transition.  At low temperatures, oscillations can be observed over the full range of measured pulse-area. As the temperature increases, the damping and the rotation period increase. %For $\Theta<6\pi$, the data can be fitted using $\Re{[K]=K_2\Omega^2},\Im{[k]}=0$. $K_2$ is proportional to temperature \cite{Ramsay_prl2010}. This is characteristic of LA-phonons, and rules out mechanisms described by activation energies such as the wetting-layer \cite{VillasBoas_prl} or LO-phonons.
 At 35~K the period of rotation is approximately 23\% larger than at 5~K.

To compare the model to the experimental data presented in fig. \ref{fig:fig2}(a), we make fits of the entire data-set to Eqs. (\ref{eqn:Bloch2},\ref{eqn:Bloch3}). This is achieved as follows. First we calculate all of the Rabi rotations using trial values for $A$ and $\hbar\omega_c$. To accommodate small variations in the dot-laser couplings between measurements at different temperatures, the calculation is scaled to the data by finding the values of $a=\Theta/\sqrt{P}$, and $\eta$ that minimize the root-mean square error for each temperature. $a$ is a measure of the dot-laser coupling, and $\eta$ is the photocurrent corresponding to one exciton. A genetic search algorithm \cite{website} is then used to find the values of $A$ and $\hbar\omega_c$ that provide the best fit of the entire data set. Note that only two parameters are used to fit the shape of the Rabi rotation, which is characterized by three qualitative features. $A$ is determined by the strength of the damping, and $\hbar\omega_c$ by the cut-off in the damping.  The nonlinearity of the rotation angle, which depends on both $A$ and $\hbar\omega_c$, provides a test of the consistency of the model.
By treating $A$ and $\hbar\omega_c$ as global fitting parameters, the temperature dependence of these features is tested against the model, and the values of $A$ and $\hbar\omega_c$ are determined with greater accuracy compared to using a series of local fits.  Good fits to the entire data-set between 5-50 K are achieved. This confirms that the weak-coupling model used here provides a good description of the experimental data. Note that $\Gamma_2^*=0.025~\mathrm{ps^{-1}}$ is not used as an adjustable fitting parameter.

Based on fits to the data in fig. \ref{fig:fig2}(a), we extract values of $A=10.8-11.9 ~\mathrm{fs.K^{-1}}$, and $\hbar\omega_c = 1.37-1.46 ~\mathrm{meV}$. The value of $A$ is consistent with a calculation using bulk GaAs material parameters \cite{Ramsay_prl2010}. %Hence,  additional contributions due to extrinsic mechanisms, such as the wetting layer, must be weak.
The cut-off energy $\hbar\omega_c$ corresponds to a spherical \cite{footnote2} wavefunction of $\psi(x)\propto e^{-x^2/2d^2}$, with $d=\sqrt{2}c_s/\omega_c=3.25-3.46 ~\mathrm{nm}$, where $c_s$ is the speed of sound. This corresponds to a probability distribution of $\sim 5.5~\mathrm{nm}$ FWHM, which is reasonable for an InAs/GaAs quantum dot.

Figure \ref{fig:fig2}(b), presents the square-root of the incident power used to reach the first minimum in the Rabi rotation $\sqrt{P}_{2\pi}$, scaled to the pulse-area at $5~\mathrm{K}$, and plotted against the temperature.
The line is calculated assuming that the scaling parameter $a$ is constant. There is good agreement between the model and the data, with a less than 9\% departure of any data point from the calculation. This implies that the increase in the period with temperature shown in fig. \ref{fig:fig2}(b) is  due to the interaction with  phonons, and that variations in the dot-laser coupling $a$ due to the thermo-mechanical stability of the alignment are small.
 %The departure could be due to thermal expansion of the sample mount. Alternatively, if the model does not fully account for the phonon-induced shift in the Rabi rotation frequency,  this would be manifest as a temperature dependence in $\sqrt{P}_{2\pi}$.
 On close inspection of the fits in fig. \ref{fig:fig2}(a) (not shown) for $T\geq 40~\mathrm{K}$, the weak-coupling model over estimates the damping reducing the quality of the fits. Since non-Markovian effects should be more influential at low temperature \cite{Krugel_prb}, we cautiously suggest that at higher temperatures,  a model which uses the renormalized Rabi frequency for calculating the rate of damping (e.g. as accomplished by the strong-coupling approach of Ref. \cite{Wurger_prb}) may provide a better description.

%The photocurrent per exciton is plotted against the left-hand axis of fig. \ref{fig:fig2}(b), and displays a $\sim 25\%$ increase over the 5-50~K %temperature range. We attribute this to thermal activation of the hole, since the detection efficiency is limited by a hole tunneling rate that is %slow compared to the 76-MHz repetition rate of the laser \cite{Kolodka_prb}.

 For a number of reasons only LA-phonons can explain the data. In a low-driving field regime ($\Theta<6\pi$), the Rabi rotations can be described by a rate of dephasing $\Gamma_2=K_2\Omega^2$, where $K_2$ is proportional to temperature \cite{Ramsay_prl2010}. This is characteristic of LA-phonons, since other mechanisms such as LO-phonons or the wetting layer would be characterized by an activation energy. The temperature gradient $A$ is consistent with a calculation based on LA-phonons using literature values for bulk GaAs \cite{Ramsay_prl2010}. The wetting layer model, as presented in ref. \cite{VillasBoas_prl}, cannot describe the cut-off behavior in the damping, or the nonlinearity of the Rabi rotation angle. Furthermore, the spectral density of the dot-environment interaction, $J(\Omega)$ can be deduced from the Rabi rotation data, and is only consistent with an LA-phonon mechanism.

To conclude, we have studied the interplay between an optically driven quantum dot exciton and a reservoir of LA-phonons using Rabi rotation measurements. The thermal and vacuum fluctuations of the phonon bath give rise to a temperature and driving-field dependent rate of dephasing and a renormalization of the Rabi energy, which are connected by a Kramers-Kronig relationship. The vacuum contribution, analogous to the photon-induced Lamb-shift observed in the spectra of atomic hydrogen, can be inferred from the fits to the data by calculating $\Im{[K]}$ at 0~K. This is presented in the fig. \ref{fig:fig2}(c).  The Lamb-shift term depends on the Rabi frequency, and has a value of $-49\pm 3~\mathrm{\mu eV}$ at the first minimum. At low Rabi frequencies this can account for up to a $4.5\pm 0.3\%$ decrease in the effective Rabi rotation frequency. At 20~K, where $k_BT\sim\hbar\omega_c$, the thermal and vacuum contributions to the shift $\Im{[K]}$ are of a similar size, indicating that for $T\ll 20~\mathrm{K}$ the shift to the Rabi rotation frequency is dominated by the vacuum contribution.  In comparison to the photon-induced Lamb-shift, here the Rabi splitting of the optically dressed states can be tuned with a laser from zero to a few meV, probing the spectral density of the exciton-phonon coupling. By contrast, optical transitions have energies in the eV range, which is large compared with the tuning range of the transition. The phonon fluctuation-induced shift to the Rabi frequency observed here contrasts with the case of quantum wells \cite{Cundiff_prl}, where a Rabi frequency renormalization results from an increase in the excitonic dipole due to many-body Coulomb effects.

%The strength of the driving field dependent dephasing extracted from Rabi rotation measurements is proportional to temperature \cite{Ramsay_prl2010}. This is characteristic of LA-phonons, and rules out mechanisms with activation energies such as the wetting-layer \cite{VillasBoas_prl} or LO-phonons. Furthermore, the $A$-parameter is consistent with bulk GaAs material parameters, therefore additional extrinsic mechanisms are note required to explain the data.

The authors thank the EPSRC (UK) EP/G001642, and the QIPIRC UK  for financial support. AN is supported by the EPSRC, and BWL by the Royal Society. We thank H.~Y.~Liu and M.~Hopkinson for sample growth; R.~Golestanian, G.~A.~Gehring and D.~P.~S.~McCutcheon for useful discussions.

%\bibliographystyle{apsrev}
%\bibliography{Ramsay_090218}% Produces the bibliography via BibTeX.

\begin{thebibliography}{1}
\bibitem{footnote1}
Uses Eq. (2) of ref. \cite{Nazir_prb} in the resonant case ($\nu=0$), where a Markov approximation is applied by taking the integrals to infinity, and assuming $\Omega$ is quasi-constant on the timescale of the integrand. This gives Eqs. (1,2) of the main-text with $D(\infty)=\Re{[K(\Omega,T)]}$, and $\Re{[\zeta_2(\infty)]}=\Im{[K(\Omega,T)]}$.
\bibitem{footnote3}
Heating by the laser is $<2~\mathrm{K}$ at $\Theta=14\pi$. %A comparison of the red-shift with ref. \cite{Ortner_prb} indicates that the sample temperature is $<10\%$ lower than on the thermometer.
\bibitem{footnote2}
We have calculated the form-factor $\mathcal{P}(\omega)$ for asymmetric dots with different sizes of electron and hole wavefunctions. In the cases of interest $\mathcal{P}(\omega)$ can be approximated by $e^{-\omega^2/2\omega_c^2}$ described by a single parameter.
\end{thebibliography}

\end{document}